\documentclass[aps,showpacs,prd,twocolumn]{revtex4}
\usepackage{amsfonts}
\usepackage{amssymb}
\usepackage{amsmath}
\usepackage{color}
\usepackage[usenames,dvipsnames]{xcolor}
\usepackage{float}
\usepackage{accents}
\usepackage{graphicx}
\usepackage{epstopdf}
\usepackage[colorlinks=true,pdfstartview=FitV,linkcolor=blue,citecolor=blue,urlcolor=blue,breaklinks=true]{hyperref}

\input{tcilatex}

\begin{document}

\title{Topological first-order vortices in a gauged $CP\left( 2\right) $\
model}
\author{R. Casana$^{1}$, M. L. Dias$^{1}$ and E. da Hora$^{2}$.}
\affiliation{$^{1}${Departamento de F\'{\i}sica, Universidade Federal do Maranh\~{a}o,}\\
65080-805, S\~{a}o Lu\'{\i}s, Maranh\~{a}o, Brazil.\\
$^{2}$Coordenadoria Interdisciplinar de Ci\^{e}ncia e Tecnologia,\\
Universidade Federal do Maranh\~{a}o, {65080-805}, S\~{a}o Lu\'{\i}s, Maranh%
\~{a}o, Brazil{.}}

\begin{abstract}
We study time-independent radially symmetric first-order solitons in a $%
CP(2) $ model interacting with an Abelian gauge field whose dynamics is
controlled by the usual Maxwell term. In this sense, we develop a consistent
first-order framework verifying the existence of a well-defined lower bound
for the corresponding energy. We saturate such a lower bound by focusing on
those solutions satisfying a particular set of coupled first-order
differential equations. We solve these equations numerically using
appropriate boundary conditions giving rise to regular structures possessing
finite-energy. We also comment the main features these configurations
exhibit. Moreover, we highlight that, despite the different solutions we
consider for an auxiliary function $\beta \left( r\right) $ labeling the
model (therefore splitting our investigation in two a priori distinct
branches), all resulting scenarios engender the very same phenomenology,
being physically equivalent.
\end{abstract}

\maketitle

\section{Introduction}

\label{Intro}

Topological objects are frequently described as the time-independent regular
solutions possessing finite-energy arising from highly nonlinear
Euler-Lagrange equations in the presence of appropriated boundary conditions 
\cite{n5}. In some particular cases, the Bogomol'nyi-Prasad-Sommerfield
(BPS) formalism allows to show that these solutions can also satisfy a set
of coupled first-order differential equations, the BPS ones \cite{n4}.

In this sense, vortices are radially symmetric solitonic configurations
appearing in a planar scalar scenario endowed by a gauge field, their
energies being commonly proportional to the magnetic flux, both ones being
quantized,\ i.e. proportional to an integer winding number. In the context
of the classical field theory, these structures were firstly studied within
the Maxwell-Higgs system, the self-interacting potential possessing no
symmetric vacuum engendering topological first-order vortex solutions only 
\cite{n1}. Similar objects were also encountered in a Chern-Simons-Higgs
scenario, the corresponding theory giving rise to both topological and
nontopological first-order vortices \cite{cshv}.

Furthermore, during the last years, many additional progress have been made
regarding first-order vortices arising within different theoretical
scenarios, including generalizations of the Abelian-Higgs theories \cite%
{gaht}, Lorentz-violating systems \cite{lvs} and gauged models with
noncanonical kinetic terms \cite{gmnckt}, many of them being applied as an
attempt to explain different cosmological and gravitational phenomena \cite%
{ames}.

In particular, in view of the developments introduced in \cite{cp1}, the
question on whether a gauged $CP(N-1)$ model with $N>2$ supports topological
vortices arises in a rather natural way, mainly due to the straight relation
between the $CP(N-1)$ theory and the four-dimensional Yang-Mills-Higgs one,
the first mapping some interesting properties of the second \cite{cpn-1}.

In a very recent contribution \cite{loginov}, the author have investigated
topological vortices arising within a gauged $CP(N-1)$ model, the
electromagnetic and the $CP(N-1)$ scalar fields interacting minimally, the
dynamics of the gauge sector being controlled by the usual Maxwell term. In
that work, the author have considered only non-BPS profiles, the
corresponding solutions being obtained via the second-order Euler-Lagrange
equations, from which the author have also studied interesting properties of
the resulting structures, such as the way the energy and magnetic flux
depend on the parameters of the overall model.

We now go a little bit further into such theme by considering the very same
theoretical scenario, but now looking for a general first-order framework
engendering radially symmetric vortices, whilst studying how these
structures differ from their non-BPS counterparts.

In order to present our results, this manuscript is organized as follows:
firstly, in the next Section II, we introduce the gauged $CP(N-1)$ model we
study, presenting some basic definitions and conventions. We particularize
our investigation by focusing on the time-independent fields giving rise to
radially symmetric configurations. In such a scenario, we look for a
consistent first-order framework by manipulating the expression for the
effective energy functional in order to establish a well-defined lower bound
for the corresponding total energy (here, it is important to point out that
such construction is only possible when a particular constraint involving
the potential is fulfilled). We verify that this bound is saturated (the
energy being minimized) when the profile functions describing the original
fields satisfy a particular set of coupled first-order equations. We also
calculate general results for the energy bound and the magnetic flux
inherent to the BPS configurations. In the Section III, we investigate in
detail the way the first-order expressions introduced previously generate
legitimate solutions, whilst using the aforecited constraint to determine
the potential term defining the vacuum manifold of the corresponding model.
Then, we solve the first-order equations numerically via convenient boundary
conditions, from which we obtain regular configurations possessing finite
energy, also identifying and commenting the main features these solitons
engender. Finally, in the Sec. IV, we present our conclusions and general
perspectives regarding future contributions.

We highlight that, throughout all this manuscript, we use the natural units
system together with $\eta ^{\mu \nu }=\left( +--\right) $ for the planar
metric signature.

\section{The overall model \label{2}}

\label{general}

We begin our investigation by considering the planar gauged $CP(N-1)$ model
introduced in \cite{loginov}, its Lagrange density reading%
\begin{equation}
\mathcal{L}=-\frac{1}{4}F_{\mu \nu }F^{\mu \nu }+\left( P_{ab}D_{\mu }\phi
_{b}\right) ^{\ast }P_{ac}D^{\mu }\phi _{c}-V\left( \left\vert \phi
\right\vert \right) \text{,}  \label{ai1}
\end{equation}%
with $F_{\mu \nu }=\partial _{\mu }A_{\nu }-\partial _{\nu }A_{\mu }$ being
the usual electromagnetic field strength tensor, $P_{ab}=\delta
_{ab}-h^{-1}\phi _{a}\phi _{b}^{\ast }$ standing for a projection operator
defined conveniently and $D_{\mu }\phi _{a}=\partial _{\mu }\phi
_{a}-igA_{\mu }Q_{ab}\phi _{b}$ representing the corresponding covariant
derivative (here, $Q_{ab}$ is a charge matrix, diagonal and real).
Additionally, the $CP(N-1)$ field $\phi $ is constrained to satisfy $\phi
_{a}^{\ast }\phi _{a}=h$. In our notation, Greek indexes run over time-space
coordinates, the Latin ones counting the complex components of the $CP(N-1)$
field themselves.

It is instructive to write down the static Gauss law coming from (\ref{ai1}%
), i.e. (here, $j$ runs over spatial coordinates only)%
\begin{equation}
\partial _{j}\partial ^{j}A^{0}=J^{0}\text{,}  \label{aig1}
\end{equation}%
the charge density being%
\begin{equation}
J^{0}=ig\left[ \left( P_{ab}D^{0}\phi _{b}\right) ^{\ast }P_{ac}Q_{cd}\phi
_{d}-P_{ab}D^{0}\phi _{b}\left( P_{ac}Q_{cd}\phi _{d}\right) ^{\ast }\right] 
\text{,}
\end{equation}%
where $D^{0}\phi _{b}=-igQ_{bc}\phi _{c}A^{0}$. In this sense, according the
Gauss law (\ref{aig1}), one concludes that the static configurations that
the model (\ref{ai1}) engenders have null electric charge, being then
compatible with the gauge condition $A^{0}=0$ which satisfies (\ref{aig1})
identically.

In this work, for the sake of simplicity, we consider the $N=3$ case, this
way reducing our study to the $CP(2)$ scenario. In such a context, we look
for time-independent solutions with radial symmetry. In this sense, we guide
our calculations using the standard map (with winding numbers $m_{1}$, $%
m_{2} $ and $m_{3}\in \mathbb{Z}$) 
\begin{equation}
A_{i}=-\frac{1}{gr}\epsilon ^{ij}n^{j}A\left( r\right) \text{,}  \label{li0}
\end{equation}%
\begin{equation}
\left( 
\begin{array}{c}
\phi _{1} \\ 
\phi _{2} \\ 
\phi _{3}%
\end{array}%
\right) =h^{\frac{1}{2}}\left( 
\begin{array}{c}
e^{im_{1}\theta }\sin \left( \alpha \left( r\right) \right) \cos \left(
\beta \left( r\right) \right) \\ 
e^{im_{2}\theta }\sin \left( \alpha \left( r\right) \right) \sin \left(
\beta \left( r\right) \right) \\ 
e^{im_{3}\theta }\cos \left( \alpha \left( r\right) \right)%
\end{array}%
\right) \text{,}  \label{li1}
\end{equation}%
with $\epsilon ^{ij}$ and $n^{j}=\left( \cos \theta ,\sin \theta \right) $
being the bidimensional Levi-Civita tensor ($\epsilon ^{12}=+1$) and the
position unit vector, respectively, the magnetic field being expressed as%
\begin{equation}
B\left( r\right) =-\frac{1}{gr}\frac{dA}{dr}\text{.}
\end{equation}

It is worthwhile to note that, once we are interested in regular solutions
presenting no divergences, the profile functions $\alpha \left( r\right) $
and $A\left( r\right) $ must obey%
\begin{equation}
\alpha (r\rightarrow 0)\rightarrow 0\text{ \ \ and \ \ }A(r\rightarrow
0)\rightarrow 0\text{.}  \label{ai00}
\end{equation}%
As we demonstrate below, we use these conditions to obtain well-behaved
field solutions related to (\ref{ai1}).

We highlight that, regarding the combination between the charge matrix $%
Q_{ab}$ and the winding numbers in (\ref{li1}), there are two possibilities
engendering topological solitons: (i) $Q=\lambda _{3}/2$ and $m_{1}=-m_{2}=m$%
, and (ii) $Q=\lambda _{8}/2$ and $m_{1}=m_{2}=m$ (both ones with $m_{3}=0$, 
$\lambda _{3}$ and $\lambda _{8}$ being the diagonal Gell-Mann matrices,
i.e. $\lambda _{3}=$diag$\left( 1,-1,0\right) $ and $\sqrt{3}\lambda _{8}=$%
diag$\left( 1,1,-2\right) $). Nevertheless, in \cite{loginov}, the author
have demonstrated that these two combinations simply mimic each other, this
way existing only one effective scenario. Therefore, in this work, we
investigate only the case defined by $m_{1}=-m_{2}=m$, $m_{3}=0$ and%
\begin{equation}
Q_{ab}=\frac{1}{2}\lambda _{3}=\frac{1}{2}\text{diag}\left( 1,-1,0\right) 
\text{,}
\end{equation}%
for convenience.

The differential equation for the profile function $\beta \left( r\right) $
reads%
\begin{equation}
\frac{d^{2}\beta }{dr^{2}}+\left( \frac{1}{r}+2\frac{d\alpha }{dr}\cot
\alpha \right) \frac{d\beta }{dr}=\frac{\sin ^{2}\alpha \sin \left( 4\beta
\right) }{r^{2}}\left( m-\frac{A}{2}\right) ^{2}\text{,}
\end{equation}%
giving rise to two constant solutions, i.e. 
\begin{equation}
\beta \left( r\right) =\beta _{1}=\frac{\pi }{4}+\frac{\pi }{2}k\text{ \ \
or \ \ }\beta \left( r\right) =\beta _{2}=\frac{\pi }{2}k\text{,}
\label{betha}
\end{equation}%
with $k\in \mathbb{Z}$. A priori, these solutions define two different
cases. However, as we demonstrate later below, these cases engender the very
same phenomenology, at least when concerning the first-order results at the
classical level.

It is important to say that, from this point on, our expressions effectively
describe the scenario defined by the choices we have introduced above.

We focus our attention on those solutions satisfying a particular set of
coupled first-order equations. In this work, we obtain these equations by
following the canonical approach, i.e. by requiring the minimization of the
total energy of the system, the starting-point being the expression for the
energy-momentum tensor related to the model (\ref{ai1}), i.e.%
\begin{equation}
\mathcal{T}_{\lambda \rho }=-F_{\lambda \mu }F_{\rho }{}^{\mu }+2\left(
P_{ab}D_{\lambda }\phi _{b}\right) ^{\ast }P_{ac}D_{\rho }\phi _{c}-\eta
_{\lambda \rho }\mathcal{L}\text{,}
\end{equation}%
from which one gets the radially symmetric energy density ($\varepsilon
\equiv \mathcal{T}_{00}=-\mathcal{L}$)%
\begin{equation}
\varepsilon =\frac{B^{2}}{2}+h\left[ \left( \frac{d\alpha }{dr}\right) ^{2}+%
\frac{W}{r^{2}}\left( \frac{A}{2}-m\right) ^{2}\sin ^{2}\alpha \right] +V%
\text{,}  \label{ai2}
\end{equation}%
the corresponding total energy reading%
\begin{eqnarray}
\frac{E}{2\pi } &=&\int \left( \frac{B^{2}}{2}+V\right) rdr  \label{te3} \\
&&\hspace{-1cm}+h\int \left[ \left( \frac{d\alpha }{dr}\right) ^{2}+\frac{W}{%
r^{2}}\left( \frac{A}{2}-m\right) ^{2}\sin ^{2}\alpha \right] rdr\text{,} 
\notag
\end{eqnarray}%
where we have introduced the auxiliary function%
\begin{equation}
W=W\left( \alpha ,\beta \right) =1-\sin ^{2}\alpha \cos ^{2}\left( 2\beta
\right) \text{,}
\end{equation}%
the solution for $\beta \left( r\right) $ being necessarily one of those
stated in (\ref{betha}).

In order to find the correspondent first-order equations, we write the
expression (\ref{te3}) in the convenient form%
\begin{eqnarray}
&&\frac{E}{2\pi }=\frac{1}{2}\int \left( B\mp \sqrt{2V}\right) ^{2}rdr
\label{te3.1} \\
&&+h\int \left[ \frac{d\alpha }{dr}\mp \frac{\sqrt{W}}{r}\left( \frac{A}{2}%
-m\right) \sin \alpha \right] ^{2}rdr  \notag \\
&&\hspace{-1.1cm}\mp \int \left[ \frac{d\left( A-2m\right) }{dr}\frac{\sqrt{%
2V}}{g}+\left( A-2m\right) h\sqrt{W}\frac{d\left( \cos \alpha \right) }{dr}%
\right] dr\text{.}  \notag
\end{eqnarray}%
The expression in the third row above can be converted in a total derivative
whether we consider the fundamental constraint%
\begin{equation}
\frac{1}{g}\frac{d}{dr}\left( \sqrt{2V}\right) =h\sqrt{W}\frac{d\left( \cos
\alpha \right) }{dr}\text{.}  \label{ai0}
\end{equation}%
This way, the energy (\ref{te3.1}) can be rewritten as%
\begin{eqnarray}
&&E=E_{bps}+\pi \int \left( B\mp \sqrt{2V}\right) ^{2}rdr  \label{te4} \\
&&\text{ \ \ \ \ \ }+2\pi h\int \left[ \frac{d\alpha }{dr}\mp \frac{\sqrt{W}%
}{r}\left( \frac{A}{2}-m\right) \sin \alpha \right] ^{2}rdr\text{,}  \notag
\end{eqnarray}%
where we have defined $E_{bps}$ as%
\begin{equation}
E_{bps}=2\pi \int r\varepsilon _{bps}dr\text{,}
\end{equation}%
with $\varepsilon _{bps}$ being given by%
\begin{equation}
\varepsilon _{bps}=\mp \frac{1}{gr}\frac{d}{dr}\left[ \left( A-2m\right) 
\sqrt{2V}\right] \text{.}
\end{equation}%
The quantity $E_{bps}$\ is finite and positive when the potential fulfills $%
V(r\rightarrow \infty )=0$, with $V_{0}\equiv V\left( r\rightarrow 0\right) $%
\ being finite and positive also.

Equation (\ref{te4}) shows that the corresponding energy exhibits a
well-defined lower bound (a property inherent to such a first-order
construction), this bound being saturated when the profile functions $\alpha
\left( r\right) $ and $A\left( r\right) $ obey%
\begin{equation}
B=\pm \sqrt{2V}\text{,}  \label{ai4}
\end{equation}%
\begin{equation}
\frac{d\alpha }{dr}=\pm \frac{\sin \alpha }{r}\left( \frac{A}{2}-m\right) 
\sqrt{1-\sin ^{2}\alpha \cos ^{2}\left( 2\beta \right) }\text{,}  \label{ai5}
\end{equation}%
which are the effective first-order equations the model (\ref{ai1})
engenders. In this sense, when (\ref{ai4}) and (\ref{ai5}) are satisfied,
the total energy of the resulting configurations can be evaluated directly,
reading (the upper (lower) sign holding for negative (positive) values of $m$%
)%
\begin{equation}
E=E_{bps}=\mp \frac{4\pi }{g}m\sqrt{2V_{0}}\text{,}  \label{ai6}
\end{equation}%
being quantized in terms of $m$ itself.

Another quantity commonly referred when investigating first-order vortices
is the magnetic flux $\Phi _{B}$ they exhibit. In the present case, the
resulting flux reads%
\begin{equation}
\Phi _{B}=2\pi \int rB\left( r\right) dr=-\frac{2\pi }{g}A_{\infty }\text{,}
\label{ai7}
\end{equation}%
where $A_{\infty }\equiv A\left( r\rightarrow \infty \right) $ must be
chosen in order to fulfill the finite-energy requirement $\varepsilon \left(
r\rightarrow \infty \right) \rightarrow 0$. Moreover, as we demonstrate
below, once $A_{\infty }$\ is specified, the energy bound $E_{bps}$\ (\ref%
{ai6}) can be proven to be proportional to the magnetic flux $\Phi _{B}$\ (%
\ref{ai7}), such relation being expected to occur during the study of
Abelian gauged first-order vortices.

In the next Section, we use the first-order expressions we have introduced
above to generate regular solutions with finite-energy. It is important to
say that the first-order equations (\ref{ai4}) and (\ref{ai5}), together
with the fundamental constraint (\ref{ai0}), solve the second-order
Euler-Lagrange ones coming from (\ref{ai1}), therefore providing genuine
solutions of the model.

%%%%%%%%%%%%%%%%%%%%%%%%

\section{The first-order solutions}

Now, we investigate the solutions the first-order framework we have
developed provides. Here, we proceed as follows: firstly, we choose one
particular solution for $\beta \left( r\right) $ coming from (\ref{betha}).
In the sequel, we use such a solution to solve the differential constraint (%
\ref{ai0}), from which we obtain the corresponding potential $V\left( \alpha
\right) $ related to that particular case. A posteriori, we implement these
both solutions (i.e. for $\beta \left( r\right) $\ and $V\left( \alpha
\right) $) into the general expression for the energy density (\ref{ai2}),
this way getting the asymptotic boundary conditions the profile functions $%
\alpha \left( r\right) $ and $A\left( r\right) $ must satisfy in order to
generate finite-energy structures. Then, using these conditions and the ones
in (\ref{ai00}), we solve the resulting first-order equations (\ref{ai4})
and (\ref{ai5}) numerically, obtaining the corresponding solutions for $%
\alpha \left( r\right) $ and $A\left( r\right) $. Finally, we depict these
solutions and the physical profiles they engender (energy density and
magnetic field), also calculating their total energies and magnetic fluxes
explicitly, from which we compare the final scenarios.

Then, let us consider the cases $\beta \left( r\right) =\beta _{1}$ and $%
\beta \left( r\right) =\beta _{2}$ separately.

\subsection{The $\protect\beta \left( r\right) =\protect\beta _{1}$ case}

This case was partially considered in \cite{loginov}, the respective author
suggesting that such construction was possible. Here, we go a little bit
further in such a investigation, showing that the general first-order
framework we have introduced recovers the expressions proposed in that work
(additionally reinforcing the coherence of our construction).

We start by choosing%
\begin{equation}
\beta \left( r\right) =\beta _{1}=\frac{\pi }{4}+\frac{\pi }{2}k\text{,}
\label{v10}
\end{equation}%
from which one gets that $\cos ^{2}\left( 2\beta _{1}\right) =0$, the
constraint (\ref{ai0}) reducing to%
\begin{equation}
\frac{d}{dr}\left( \sqrt{2V}\right) =\frac{d}{dr}\left( gh\cos \alpha
\right) \text{,}
\end{equation}%
whose solution is (we have used $C=0$ for the integration constant)%
\begin{equation}
V\left( \alpha \right) =\frac{g^{2}h^{2}}{2}\cos ^{2}\alpha \text{,}
\label{v1}
\end{equation}%
i.e. the potential related to the $\beta \left( r\right) =\beta _{1}$ case,
the resulting first-order equations (\ref{ai4}) and (\ref{ai5}) standing for%
\begin{equation}
\frac{1}{r}\frac{dA}{dr}=\mp g^{2}h\cos \alpha \text{,}  \label{v12}
\end{equation}%
\begin{equation}
\frac{d\alpha }{dr}=\pm \frac{\sin \alpha }{r}\left( \frac{A}{2}-m\right) 
\text{.}  \label{v13}
\end{equation}

Now, in order to solve the equations (\ref{v12}) and (\ref{v13}), beyond the
behavior in (\ref{ai00}), we need to specify the conditions the profile
functions $\alpha \left( r\right) $ and $A\left( r\right) $ obey in the
asymptotic limit (these conditions ensuring that the total energy of the
final configurations is finite). In this sense, we implement (\ref{v10}) and
(\ref{v1}) into the basic expression (\ref{ai2}), from which we get%
\begin{equation}
\varepsilon =\frac{B^{2}}{2}+h\left[ \left( \frac{d\alpha }{dr}\right) ^{2}+%
\frac{W_{1}}{r^{2}}\left( \frac{A}{2}-m\right) ^{2}\right] +\frac{g^{2}h^{2}%
}{2}\cos ^{2}\alpha \text{,}
\end{equation}%
with%
\begin{equation}
W_{1}\equiv W\left( \alpha ,\beta =\beta _{1}\right) =\sin ^{2}\alpha \text{,%
}
\end{equation}%
the corresponding energy being finite for $\varepsilon \left( r\rightarrow
\infty \right) \rightarrow 0$, $\alpha \left( r\right) $ and $A\left(
r\right) $ being then constrained to satisfy%
\begin{equation}
\alpha \left( r\rightarrow \infty \right) \rightarrow \frac{\pi }{2}\text{ \
\ and \ \ }A\left( r\rightarrow \infty \right) \rightarrow 2m\text{,}
\label{v14}
\end{equation}%
with $m\in\mathbb{Z}$.

At this time, one can easily calculate the total energy $E_{bps}$ (\ref{ai6}%
) and the magnetic flux $\Phi _{B}$ (\ref{ai7}) of the resulting solitons,
i.e.%
\begin{equation}
E_{bps}=\mp 4\pi hm\text{ \ \ and \ \ }\Phi _{B}=-\frac{4\pi }{g}m\text{,}
\end{equation}%
(with $E_{bps}=\pm gh\Phi _{B}$) both ones being quantized in terms of the
winding number $m$, as expected.

It is also interesting to investigate the way the fields $\alpha \left(
r\right) $ and $A\left( r\right) $ approximate the boundary values in (\ref%
{ai00}) and (\ref{v14}). In order to perform such calculation, in what
follows, we consider $m>0$ only (i.e. the lower signs in the first-order
expressions), for simplicity. Then, following the usual algorithm, we
linearize the equations (\ref{v12}) and (\ref{v13}) in order to get the
approximate solutions near the origin (with $\lambda =g^{2}h/2$)%
\begin{equation}
\alpha \left( r\right) \approx C_{1}r^{m}\text{ \ \ and \ \ }A\left(
r\right) \approx \lambda r^{2}\text{,}  \label{v15}
\end{equation}%
and in the asymptotic regime%
\begin{equation}
\alpha \left( r\right) \approx \frac{\pi }{2}-C_{2}e^{-M_{\alpha }r}\text{ \
\ and \ \ }A\left( r\right) \approx 2m-2C_{2}\sqrt{\lambda }re^{-M_{A}r}%
\text{,}  \label{v16}
\end{equation}%
$M_{\alpha }=M_{A}=\sqrt{\lambda }$ being the masses of the corresponding
bosons (the relation $M_{\alpha }/M_{A}=1$ typically defining the
Bogomol'nyi limit), $C_{1}$ and $C_{2}$ standing for positive constants to
be fixed by requiring the correct behavior near the origin and
asymptotically.

We summarize the scenario as follows: given the profile functions $\alpha
\left( r\right) $ and $A\left( r\right) $ satisfying the equations (\ref{v12}%
) and (\ref{v13}), and obeying the boundary conditions (\ref{ai00}) and (\ref%
{v14}), the resulting radially symmetric first-order vortices exhibit
quantized energy and magnetic flux given by $E_{bps}=\mp 4\pi hm$ and $\Phi
_{B}=-4\pi m/g$, respectively, whilst approaching the boundaries according
the approximate solutions in (\ref{v15}) and (\ref{v16}).

It is interesting to note that, whether we implement $\lambda =g^{2}h/2$,
the potential in (\ref{v1}) can be written as $V\left( \left\vert \phi
_{3}\right\vert \right) =\lambda \left\vert \phi _{3}\right\vert ^{2}$,
mimicking exactly the Eq. (15) in \cite{loginov}. Moreover, in that same
work, the author suggested that the choice $\beta \left( r\right) =\beta
_{2} $ does not support first-order solutions. Nevertheless, as we will
demonstrate below, once we choose the potential conveniently, the case with $%
\beta \left( r\right) =\beta _{2}$ indeed admits coherent first-order
solitons.

\subsection{The $\protect\beta \left( r\right) =\protect\beta _{2}$ case}

Now, we go further in our investigation by choosing%
\begin{equation}
\beta \left( r\right) =\beta _{2}=\frac{\pi }{2}k\text{,}  \label{carai}
\end{equation}%
whilst giving rise to $\cos ^{2}\left( 2\beta _{2}\right) =1$, the
corresponding constraint being%
\begin{equation}
\frac{d}{dr}\left( \sqrt{2V}\right) =\frac{d}{dr}\left( \frac{gh}{2}\cos
^{2}\alpha \right) \text{,}
\end{equation}%
engendering the solution%
\begin{equation}
V\left( \alpha \right) =\frac{g^{2}h^{2}}{32}\cos ^{2}\left( 2\alpha \right) 
\text{,}  \label{v17}
\end{equation}%
which stands for the potential related to $\beta \left( r\right) =\beta _{2}$%
. Here, we have used $\mathcal{C}=-gh/4$ for the integration constant.

In this case, the first-order equations read%
\begin{equation}
\frac{1}{r}\frac{dA}{dr}=\mp \frac{g^{2}h}{4}\cos \left( 2\alpha \right) 
\text{,}  \label{2a}
\end{equation}%
\begin{equation}
\frac{d\alpha }{dr}=\pm \frac{\sin \left( 2\alpha \right) }{2r}\left( \frac{A%
}{2}-m\right) \text{.}  \label{2b}
\end{equation}

In order to determine the conditions the fields $\alpha \left( r\right) $
and $A\left( r\right) $ satisfy in the asymptotic limit, we proceed as
before, i.e. given (\ref{carai}) and (\ref{v17}), the expression (\ref{ai2})
for the energy density reduces to%
\begin{equation}
\varepsilon =\frac{B^{2}}{2}+h\left[ \left( \frac{d\alpha }{dr}\right) ^{2}+%
\frac{W_{2}}{r^{2}}\left( \frac{A}{2}-m\right) ^{2}\right] +\frac{g^{2}h^{2}%
}{32}\cos ^{2}\left( 2\alpha \right) \text{,}
\end{equation}%
where%
\begin{equation}
W_{2}\equiv W\left( \alpha ,\beta =\beta _{2}\right) =\sin ^{2}\alpha \cos
^{2}\alpha \text{.}
\end{equation}%
\begin{figure}[tbp]
\centering\includegraphics[width=8.5cm]{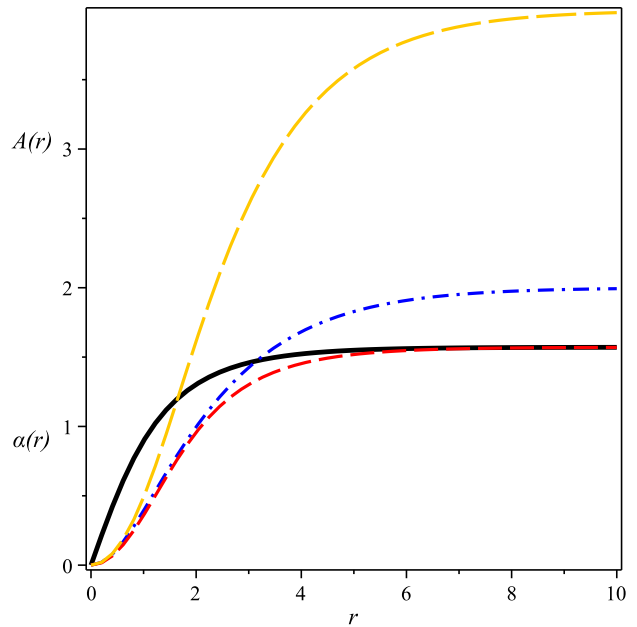}
\par
\vspace{-0.3cm}
\caption{Numerical solutions to $\protect\alpha \left( r\right) $ (solid
black line for $m=1$ and dashed red line for $m=2$) and $A\left( r\right) $
(dot-dashed blue line for $m=1$ and long-dashed orange line for $m=2$)
obtained via the first-order equations (\protect\ref{v12}) and (\protect\ref%
{v13}) in the presence of the boundary conditions (\protect\ref{ai00}) and (%
\protect\ref{v14}), for $g=h=1$.}
\end{figure}

Then, in order to fulfill the finite-energy requirement $\varepsilon \left(
r\rightarrow \infty \right) \rightarrow 0$, $\alpha \left( r\right) $ and $%
A\left( r\right) $ must behave as%
\begin{equation}
\alpha \left( r\rightarrow \infty \right) \rightarrow \frac{\pi }{4}\text{ \
\ and \ \ }A\left( r\rightarrow \infty \right) \rightarrow 2m\text{,}
\label{xx1}
\end{equation}%
from which we can also calculate the energy $E_{bps}$ and the magnetic flux $%
\Phi _{B}$ inherent to the resulting structures, i.e.%
\begin{equation}
E_{bps}=\mp \pi hm\text{ \ \ and \ \ }\Phi _{B}=-\frac{4\pi }{g}m\text{,}
\end{equation}%
($E_{bps}=\pm gh\Phi _{B}/4$) which are again quantized in terms of $m$.

Furthermore, we look for the way $\alpha \left( r\right) $ and $A\left(
r\right) $ behave near the boundaries by linearizing the first-order
equations (\ref{2a}) and (\ref{2b}) around the contours values in (\ref{ai00}%
) and (\ref{xx1}) (again, for simplicity, we use $m>0$), from which one gets
the approximate profiles near the origin%
\begin{equation}
\alpha \left( r\right) \approx \mathcal{C}_{1}r^{m}\text{ \ \ and \ \ }%
A\left( r\right) \approx \frac{\lambda }{4}r^{2}\text{,}  \label{xx2}
\end{equation}%
and in the limit $r\rightarrow \infty $, i.e.%
\begin{equation}
\alpha \left( r\right) \approx \frac{\pi }{4}-\mathcal{C}_{2}e^{-\mathcal{M}%
_{\alpha }r}\text{ \ \ and \ \ }A\left( r\right) \approx 2m-2\mathcal{C}_{2}%
\sqrt{\lambda }re^{-\mathcal{M}_{A}r}\text{,}  \label{xx3}
\end{equation}%
with $\mathcal{M}_{\alpha }=\mathcal{M}_{A}=\sqrt{\lambda }/2$ standing for
the masses of the related particles ($\mathcal{M}_{\alpha }/\mathcal{M}%
_{A}=1 $ still holding), $\mathcal{C}_{1}$ and $\mathcal{C}_{2}$ being
positive constants to be fixed via the same manner as before.

In such a scenario, given the conditions (\ref{ai00}) and (\ref{xx1}), the
solutions $\alpha \left( r\right) $ and $A\left( r\right) $ that the
equations (\ref{2a}) and (\ref{2b}) provide generate first-order vortices
possessing total energy $E_{bps}=\mp \pi hm$ and magnetic flux $\Phi
_{B}=-4\pi m/g$, behaving as the approximate profiles (\ref{xx2}) and (\ref%
{xx3}) in the appropriate limit.

It is interesting to note that the potential in (\ref{v1}) can be written as
the one in (\ref{v17}) whether we implement the redefinitions $\alpha
\rightarrow 2\alpha $, $\lambda \rightarrow \lambda /4$ and $h\rightarrow
h/4 $. Notwithstanding, also the first-order equations (\ref{v12}) and (\ref%
{v13}) reduce to those in (\ref{2a}) and (\ref{2b}) following the very same
way (the corresponding energies behaving in a similar manner, the magnetic
fluxes being automatically the same). We highlight that such redefinitions
are completely compatible also at level of the second-order Euler-Lagrange
equations.

In this sense, we argue that the $\beta \left( r\right) =\beta _{2}$ case
mimics those results obtained via $\beta \left( r\right) =\beta _{1}$, both
scenarios effectively describing the same phenomenology, at least concerning
the first-order solitons at the classical level. 
\begin{figure}[tbp]
\centering\includegraphics[width=8.5cm]{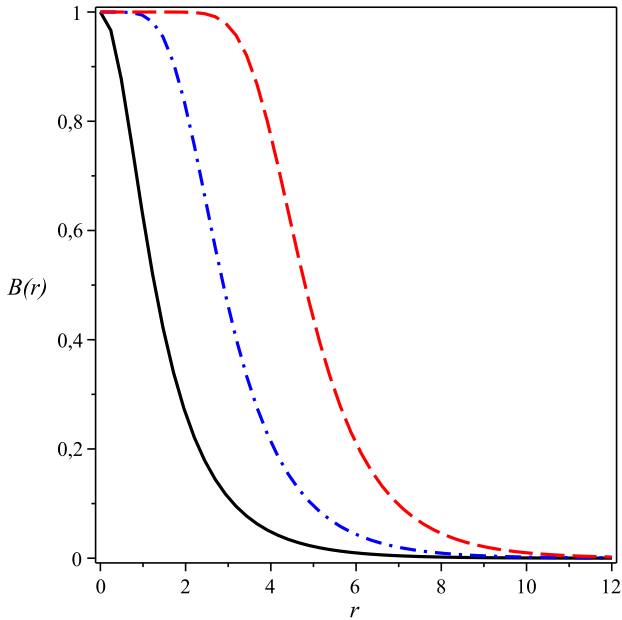}
\par
\vspace{-0.3cm}
\caption{Numerical solutions to the magnetic field $B\left( r\right) $ for $%
m=1$ (solid black line), $m=3$ (dot-dashed blue line) and $m=7$ (dashed red
line). The profiles are lumps centered at the origin.}
\end{figure}

We end this Section by presenting the numerical solutions we have found for $%
\alpha \left( r\right) $, $A\left( r\right) $, $B\left( r\right) $ and $%
\varepsilon _{bps}\left( r\right) $ via the first-order equations (\ref{v12}%
) and (\ref{v13}) in the presence of the boundary conditions (\ref{ai00})
and (\ref{v14}). We have used $g=h=1$, for simplicity. 
\begin{figure}[tbp]
\centering\includegraphics[width=8.5cm]{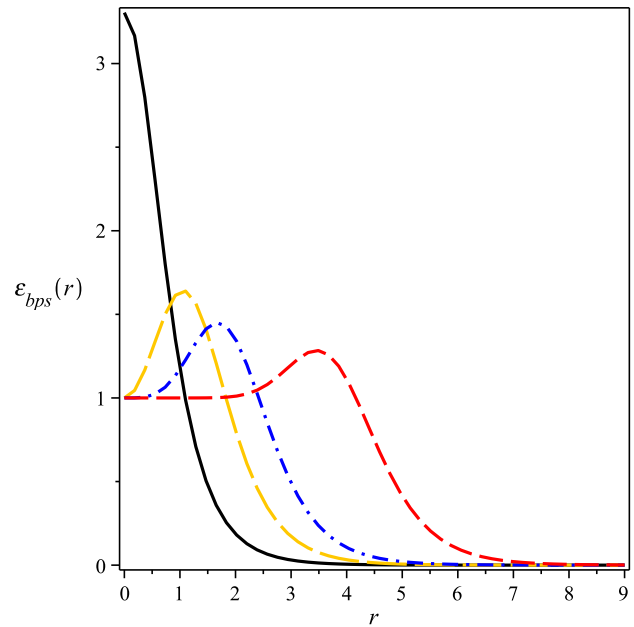}
\par
\vspace{-0.3cm}
\caption{Numerical solutions to the energy density $\protect\varepsilon %
_{bps}\left( r\right) $, conventions being the same as in the previous
Figure 2 (with the long-dashed orange line for $m=2$): for $m=1$, the
solution is a lump; for $m\neq 1$, the profiles are rings.}
\end{figure}

In the Figure 1, we depict the numerical solutions to the profile functions $%
\alpha \left( r\right) $ (solid black line for $m=1$ and dashed red line for 
$m=2$) and $A\left( r\right) $ (dot-dashed blue line for $m=1$ and
long-dashed orange line for $m=2$). In general, these profiles reach the
boundary values in a monotonic manner, behaving according the approximate
solutions we have calculated previously via the linearization of the
first-order equations. We also point out the way the gauge field coherently
fulfills the condition $A\left( r\rightarrow \infty \right) \rightarrow 2m$,
as expected.

We plot the solutions to the magnetic field $B\left( r\right) $ in the Fig.
2, for $m=1$ (solid black line), $m=3$ (dot-dashed blue line) and $m=7$
(dashed red line), all these profiles being lumps centered at the origin
(the absolute value does not depending on the winding number $m$, being
equal to the unity). Also, as the vorticity increases, the solutions spread
over greater distances, the corresponding bosons mediating large-range
interactions.

Finally, the Fig. 3 shows the profiles to the energy density $\varepsilon
_{bps}\left( r\right) $, the conventions being the same ones used in the
Figure 2. Here, for $m=1$, the resulting solution is a lump centered at the
origin. On the other hand, for $m\neq 1$, the corresponding profiles are
rings, their absolute values (amplitudes) lying on a finite distance $R$
from the origin (this way defining the "radius" of the ring). In particular,
these amplitudes (radii) decrease (increase) as the vorticity increases.

%%%%%%%%%%%%%%%%%%%%%%%%

\section{\textbf{Conclusions and perspectives}}

In this work, we have considered the gauged $CP(2)$ model proposed in \cite%
{loginov}, whilst obtaining in a coherent way the first-order vortices such
model supports.

Firstly, we have detailed the planar Lagrange density defining the overall
model, from which we have verified that such a system supports static
solutions presenting no electric field ($A^{0}=0$ satisfying the
corresponding Gauss Law identically). In the sequel, focusing our attention
on those configurations presenting radial symmetry, we have implemented the
well-known ansatz (\ref{li0}) and (\ref{li1}) together with convenient
choices for the charges and winding numbers inherent to such maps. We have
verified the constant solutions the profile function $\beta \left( r\right) $
supports, this way splitting our investigation in a priori two different
branches. In addition, whilst using a very fundamental constraint, we have
rewritten the radially symmetric expression for the effective energy
functional in such a way highlighting the existence of a lower bound for the
corresponding total energy. We have checked that the aforementioned bound is
saturated when the functions $\alpha \left( r\right) $ and $A\left( r\right) 
$ satisfy a particular set of coupled first-order equations, also
calculating the symbolic values for the total energy and the magnetic flux
the final structures exhibit.

We have divided our study according the two solutions $\beta \left( r\right) 
$ admits. Then, whilst considering these two cases separately, we have used
the fundamental constraint introduced before to obtain the particular
potentials such scenarios engender, also getting the corresponding
first-order equations and the appropriate boundary conditions. In view of
such conditions, we have calculated explicitly the values for the energy and
the magnetic flux inherent to the resulting structures. Moreover, we have
concluded that the two solutions $\beta \left( r\right) $ supports in fact
engender the very same phenomenology, being physically equivalent. Finally,
we have plotted the numerical solutions we found for the relevant profiles,
from which we have commented the general aspects they present.

It is important to emphasize that the results we have introduced in this
letter hold a priori only for those radially symmetric time-independent
configurations defined by the map (\ref{li0}) and (\ref{li1}). Therefore, it
is not possible to assure that the gauged model (\ref{ai1}) coherently
supports a first-order framework outside that map, such question lying
beyond the scope of this manuscript.

In this sense, rather natural ideas regarding new works include the search
for the nontopological first-order vortices the theoretical model we have
studied here possibly engenders. Furthermore, it is also interesting to
consider the gauged $CP(2)$ theory when endowed by the Chern-Simons term
(instead of the Maxwell one). These two possibilities are now under
investigation, and we hope positive results for a future contribution.

\begin{acknowledgments}
The authors thank CAPES, CNPq and FAPEMA (Brazilian agencies) for partial
financial support.
\end{acknowledgments}


\begin{thebibliography}{99}
\bibitem{n5} N. Manton and P. Sutcliffe, \textit{Topological Solitons}
(Cambridge University Press, Cambridge, England, 2004).

\bibitem{n4} E. Bogomol'nyi, Sov. J. Nucl. Phys. \textbf{24}, 449 (1976). M.
Prasad and C. Sommerfield, Phys. Rev. Lett. \textbf{35}, 760 (1975).

\bibitem{n1} H. B. Nielsen and P. Olesen, Nucl. Phys. B \textbf{61}, 45
(1973).

\bibitem{cshv} R. Jackiw and E. J. Weinberg, Phys. Rev. Lett. \textbf{64},
2234 (1990). R. Jackiw, K. Lee and E. J. Weinberg, Phys. Rev. D \textbf{42},
3488 (1990).

\bibitem{gaht} D. Bazeia, E. da Hora, C. dos Santos and R. Menezes, Phys.
Rev. D \textbf{81}, 125014 (2010); Eur. Phys. J. C \textbf{71}, 1833 (2011).
D. Bazeia, R. Casana, M. M. Ferreira Jr. and E. da Hora, Europhys. Lett. 
\textbf{109}, 21001 (2015). R. Casana, E. da Hora, D. Rubiera-Garcia and C.
dos Santos, Eur. Phys. J. C \textbf{75}, 380 (2015).

\bibitem{lvs} R. Casana, M. M. Ferreira Jr., E. da Hora and C. Miller, Phys.
Lett. B \textbf{718}, 620 (2012). R. Casana, M. M. Ferreira Jr., E. da Hora
and A. B. F. Neves, Eur. Phys. J. C \textbf{74}, 3064 (2014). R. Casana and
G. Lazar, Phys. Rev. D \textbf{90}, 065007 (2014). R. Casana, C. F. Farias
and M. M. Ferreira Jr., Phys. Rev. D \textbf{92}, 125024 (2015). R. Casana,
C. F. Farias, M. M. Ferreira Jr. and G. Lazar, Phys. Rev. D \textbf{94},
065036 (2016).

\bibitem{gmnckt} L. Sourrouille, Phys. Rev. D \textbf{87}, 067701 (2013). R.
Casana and L. Sourrouille, Mod. Phys. Lett. A \textbf{29}, 1450124 (2014).

\bibitem{ames} C. Armendariz-Picon, T. Damour and V. Mukhanov, Phys. Lett. B 
\textbf{458}, 209 (1999). V. Mukhanov and A. Vikman, J. Cosmol. Astropart.
Phys. \textbf{02}, 004 (2005). A. Sen, J. High Energy Phys. \textbf{07}, 065
(2002). C. Armendariz-Picon and E. A. Lim, J. Cosmol. Astropart. Phys. 
\textbf{08}, 007 (2005). J. Garriga and V. Mukhanov, Phys. Lett. B \textbf{%
458}, 219 (1999). R. J. Scherrer, Phys. Rev. Lett. \textbf{93}, 011301
(2004). A. D. Rendall, Class. Quantum Grav. \textbf{23}, 1557 (2006).

\bibitem{cp1} M. A. Mehta, J. A. Davis and I. J. R. Aitchison, Phys. Lett. B 
\textbf{281}, 86 (1992). B. M. A. G. Piette, D. H. Tchrakian and W. J.
Zakrzewski, Phys. Lett. B \textbf{339}, 95 (1994). D. H. Tchrakian and K.
Arthur, Phys. Lett. B \textbf{352}, 327 (1995). L. Sourrouille, A. Caso and
G. S. Lozano, Mod. Phys. Lett. A \textbf{26}, 637 (2011). L. Sourrouille,
Mod. Phys. Lett. A \textbf{26}, 2523 (2011); Mod. Phys. Lett. A \textbf{27},
1250094 (2012).

\bibitem{cpn-1} A. D'Adda, M. Luscher and P. D. Vecchia, Nucl. Phys. B 
\textbf{146}, 63 (1978). E. Witten, Nucl. Phys. B \textbf{149}, 285 (1979).
A. M. Polyakov, Phys. Lett. B \textbf{59}, 79 (1975). M. Shifman and A.
Yung, Rev. Mod. Phys. \textbf{79}, 1139 (2007).

\bibitem{loginov} A. Yu. Loginov, Phys. Rev. D \textbf{93}, 065009 (2016).
\end{thebibliography}
\end{document}